\documentclass[aps,prl,reprint,showpacs,twoside]{revtex4-1}

\usepackage{mathrsfs}
\usepackage{bm}
\usepackage{amsfonts}
\usepackage{amsmath}
\usepackage{amssymb}
\usepackage{graphicx}
\usepackage[colorlinks=true,linkcolor=blue,citecolor=blue,breaklinks,linktoc=page,urlcolor=blue]{hyperref}

\begin{document}
\title{Metric-Like Formulation Of the Spin-Three Gravity In Three Dimensions}

\author{Zhi-Qiang Guo}
\email{zhiqiang.guo@usm.cl}
\affiliation{Departamento de F\'{i}sica y Centro Cient\'{i}fico
Tecnol\'{o}gico de Valpara\'{i}so,\\ Universidad T\'{e}cnica Federico
Santa Mar\'{i}a,  Casilla 110-V, Valpara\'{i}so, Chile}

\begin{abstract}
We provide a metric-like formulation of the spin-3 gravity in three dimensions. It is shown that the Chern-Simons formulation of the spin-3 gravity can be reformulated as a Einstein-Cartan-Sciama-Kibble theory coupled with the higher-spin matter fields. A duality-like transformation is also identified from this metric-like formulation.
\end{abstract}


\maketitle

{\textit {Intrduction}}.{\textemdash}In three dimensions~(3D), the pure Einstein gravity does not have local degrees of freedom~\cite{Deser:1983tn}. The Einstein-Hilbert action with the cosmological constant term can be recast as a $SL(2,R){\times}SL(2,R)$ Chern-Simons~(CS) theory~\cite{Achucarro:1987vz,Witten:1988hc}, which is a manifestly topological theory. Recently, it was suggested that the higher-spin gravity~\cite{Vasiliev:1990en,Blencowe:1988gj} in 3D could also be expressed as a CS theory~\cite{Henneaux:2010xg,Campoleoni:2010zq} but with the larger gauge group $SL(N,R){\times}SL(N,R)$. In contrast with its concise formulation in terms of the frame-like fields, which facilitates the analysis of asymptotical symmetries~\cite{Brown:1986nw,Henneaux:2010xg,Campoleoni:2010zq} and higher-spin black hole solutions~\cite{Gutperle:2011kf}, a metric-like formulation of the higher-spin gravity are helpful to illuminate its geometrical structure and make its higher-spin freedoms transparent. From the perspective of anti-de Sitter/conformal field theory correspondence, the metric-like formulation in 3D is also useful to understand the thermodynamical properties~(such as entropy~\cite{Campoleoni:2012hp} and shear viscosity~\cite{Policastro:2001yc})~of its dual theory in two dimensions~\cite{Gaberdiel:2012uj}. However, a metric-like formulation can not be derived straightforwardly. A perturbative study of the metric-like formulation of the spin-3 gravity has been pursued in~\cite{Campoleoni:2012hp}. Geometrical analysis based on the metric compatibility method~\cite{Fujisawa:2012dk} shows that a complete metric-like formulation not only depends on the spin-2 field and the spin-3 field, but also higher-spin fields with more space-time indices are required. In this paper, we propose that if we assume the connection of the conventional spin-2 gravity has a torsion, then the CS formulation of the spin-3 gravity can be recast as a Einstein-Cartan-Sciama-Kibble theory~(ECSK)~\cite{Hehl:1976kj}, in which the spin-3 field can be regarded as the higher-spin matter acting as the source of the torsion.

{\textit {Metric-Like Formulation}}.{\textemdash}Similar to its spin-2 cousin in three dimensions, the spin-3 gravity in 3D could be described by the  $SL(3,R){\times}SL(3,R)$ Chern-Simons theory
\begin{eqnarray}
\label{sec2-cs-lag}
S&=&S_{\mathrm{CS}}[A]-S_{\mathrm{CS}}[\bar{A}],\\
S_{\mathrm{CS}}[A]&=&\frac{k}{4\pi}\int \mathrm{tr}\bigr(A\wedge {dA}+\frac{2}{3}A\wedge {A}\wedge {A}\bigr),\nonumber
\end{eqnarray}
where $k=\frac{l}{16G}$. $A$ and $\bar{A}$ can further be decomposed into the frame-like fields
\begin{eqnarray}
\label{sec2-cs-frame}
A=\omega+\frac{1}{l}e,~~\bar{A}=\omega-\frac{1}{l}e.
\end{eqnarray}
Then the CS action~(\ref{sec2-cs-lag}) has the Palatini formulation
\begin{eqnarray}
\label{sec2-cs-lag-pal}
S=\frac{k}{\pi}\int \mathrm{tr}\bigr(e\wedge (d\omega+\omega\wedge\omega)+\frac{1}{3l^2}e\wedge {e}\wedge {e}\bigr),
\end{eqnarray}
where the second term is the generalized cosmological term. The variation of $\omega$ yields the torsion constraints
\begin{eqnarray}
\label{sec2-cs-lag-pal-tor}
\mathscr{T}=de+\omega\wedge{e}+e\wedge\omega=0,
\end{eqnarray}
and the variation of $e$ yields the equations of motion
\begin{eqnarray}
\label{sec2-cs-lag-pal-eom}
\mathscr{R}=d\omega+\omega\wedge\omega+\frac{1}{l^2}e\wedge{e}=0.
\end{eqnarray}
If we can solve $\omega$ in terms of $e$ and $de$ through the torsion equation~(\ref{sec2-cs-lag-pal-tor}), then a second-order formulation of the Palatini action can be obtained. For the Einstein gravity, $\omega$ and $e$ take values in the Lie algebra of $SL(2,R)$, Eq.~(\ref{sec2-cs-lag-pal-tor}) can be solved straightforwardly, and a pure metric-like formulation can be achieved. For the spin-3 gravity, the Lie algebra of $\omega$ and $e$ is $SL(3,R)$. A perturbative solution of Eq.~(\ref{sec2-cs-lag-pal-tor}) has been given in~\cite{Campoleoni:2012hp}. An non-perturbative attempt has been made in~\cite{Fujisawa:2012dk}, which shows it is difficult to
achieve a pure metric-like formulation of the CS action~(\ref{sec2-cs-lag}). The work of~\cite{Fujisawa:2012dk} is based on the $SL(3,R)$ invariant metric variables $\varphi_{\alpha\beta}=\mathrm{tr}(e_{\alpha}e_{\beta})$ and $\varphi_{\alpha\beta\gamma}=\mathrm{tr}(e_{\alpha}e_{\beta}e_{\gamma})$. Alternatively, in this paper, we use the $SL(2,R)$ decomposition of the $SL(3,R)$ Lie algebra
\begin{eqnarray}
\label{sec2-lie}
[J_a,J_b]&=&\epsilon_{abc}J^{c},~~[J_a,Q_{bc}]=\epsilon_{\hspace{1mm}ab}^{d}Q_{dc}+\epsilon_{\hspace{1mm}ac}^{d}Q_{db},\nonumber\\
~[Q_{ab},Q_{cd}]&=&\lambda^2
(\eta_{ac}\epsilon_{bdm}+\eta_{bc}\epsilon_{adm})J^{m}+(c\leftrightarrow{d}),
\end{eqnarray}
where the small Latin letters take the values $0,1,2$, and the definitions of $\eta_{ab}$ and $\epsilon_{abc}$ follow the conventions in~\cite{Campoleoni:2010zq}. $\lambda$ is a dimensionless constant. The anti-commutators of $J_a$ furnish the Lie algebra of the $SL(2,R)$ group. $Q_{ab}$ is symmetrical about its indices and satisfies the traceless condition $Q_{ab}\eta^{ab}=0$. They transform as the 5D symmetrical representation of $SL(2,R)$. Using this realization of the $SL(3,R)$ Lie algebra, $\omega$ and $e$ are expressed as
\begin{eqnarray}
\label{sec2-lie-we}
\omega=\omega^{a}J_a+\omega^{bc}Q_{bc},~~e=e^{a}J_a+e^{bc}Q_{bc}.
\end{eqnarray}
Here $\omega^{bc}$ and $e^{bc}$ are also symmetrical and traceless, that is, $\omega^{bc}\eta_{bc}=0$ and $e^{bc}\eta_{bc}=0$. We define the metric-like fields from the frame-like fields as
\begin{eqnarray}
\label{sec2-lie-we-metric}
g_{\alpha\beta}=e_{\alpha}^{a}e_{\beta}^{b}\eta_{ab},~~
h_{\alpha\beta\gamma}=e_{\alpha}^{ab}e_{\beta}^{c}e_{\gamma}^{d}\eta_{ac}\eta_{bd}.
\end{eqnarray}
$g_{\alpha\beta}$ is the conventional $SL(2,R)$ invariant metric. $h_{\mu\alpha\beta}$ is only symmetrical about $\alpha$ and $\beta$, which belongs to the class of the mixed-symmetrical field discussed in~\cite{Campoleoni:2012th}. Using $g^{\alpha\beta}$ as the inverse of $g_{\alpha\beta}$, $h_{\mu\alpha\beta}$ satisfies the traceless condition $h_{\mu\alpha\beta}g^{\alpha\beta}=0$. We also have
\begin{eqnarray}
\label{sec2-lie-we-metric-anti}
e_{\alpha}^{a}e_{\beta}^{b}e_{\gamma}^{c}\epsilon_{abc}=\varepsilon_{\alpha\beta\gamma},~~
E^{\alpha}_{a}E^{\beta}_{b}E^{\gamma}_{c}\epsilon^{abc}=\varepsilon^{\alpha\beta\gamma},
\end{eqnarray}
where $g$ is the determinant of $g_{\alpha\beta}$, and $E^{\alpha}_{a}$ is the inverse of $e_{\alpha}^{a}$, which satisfies $E^{\alpha}_{a}e_{\alpha}^{b}=\delta^{a}_{b}$ and $E^{\alpha}_{a}e_{\beta}^{a}=\delta^{\alpha}_{\beta}$. We have defined
\begin{eqnarray}
\label{sec2-lie-we-metric-anti-def}
\varepsilon_{\alpha\beta\gamma}=\sqrt{-g}\epsilon_{\alpha\beta\gamma},~~
\varepsilon^{\alpha\beta\gamma}=\frac{1}{\sqrt{-g}}\epsilon^{\alpha\beta\gamma},
\end{eqnarray}
which are covariant antisymmetrical tensors under the general 3D coordinate transformation. By means of the $SL(2,R)$ variables, the torsion equation can be rewritten as
\begin{subequations}
\begin{eqnarray}
\label{sec2-cs-lag-re-tor-1}
(\partial_{\mu}e_{\nu}^{a}
&+&\omega_{\mu}^{b}e_{\nu}^{c}\epsilon_{\hspace{1mm}bc}^{a})-(\partial_{\nu}e_{\mu}^{a}
+\omega_{\nu}^{b}e_{\mu}^{c}\epsilon_{\hspace{1mm}bc}^{a})\\
&=&
4\lambda^2(\omega_{\nu{d}}^{b}e_{\mu}^{dc}\epsilon_{\hspace{1mm}bc}^{a}-
\omega_{\mu{d}}^{b}e_{\nu}^{dc}\epsilon_{\hspace{1mm}bc}^{a}),\nonumber\\
\label{sec2-cs-lag-re-tor-2}
(\partial_{\mu}e_{\nu}^{bc}&+&\omega_{\mu}^{a}e_{\nu}^{dc}\epsilon_{\hspace{1mm}ad}^{b}
+\omega_{\mu}^{a}e_{\nu}^{db}\epsilon_{\hspace{1mm}ad}^{c})-(\mu\leftrightarrow\nu)\\
&=&(\omega_{\nu}^{ab}e_{\mu}^{d}\epsilon_{\hspace{1mm}ad}^{c}
+\omega_{\nu}^{ac}e_{\mu}^{d}\epsilon_{\hspace{1mm}ad}^{b})-(\mu\leftrightarrow\nu).\nonumber
\end{eqnarray}
\end{subequations}
Eqs.~(\ref{sec2-cs-lag-re-tor-1}) and~(\ref{sec2-cs-lag-re-tor-2}) have clear interpretations in term of the $SL(2,R)$ variables. The left side of Eq.~(\ref{sec2-cs-lag-re-tor-1}) can be interpreted as the torsion of the $SL(2,R)$ frame-like fields $e_{\nu}^{a}$. The left side of Eq.~(\ref{sec2-cs-lag-re-tor-2}) transforms as a symmetrical representation of the the $SL(2,R)$ group. These observations provide us with the hints that Eqs.~(\ref{sec2-cs-lag-re-tor-1}) and~(\ref{sec2-cs-lag-re-tor-2}) can be reformulated as equations of metric-like fields through the assumptions
\begin{eqnarray}
\label{sec2-cs-tor-assm-1}
\partial_{\mu}e_{\nu}^{a}
+\omega_{\mu}^{b}e_{\nu}^{c}\epsilon_{\hspace{1mm}bc}^{a}=\Gamma^{\rho}_{\mu\nu}e_{\rho}^{a}
\end{eqnarray}
and
\begin{eqnarray}
\label{sec2-cs-tor-assm-2}
\omega_{\mu}^{bc}=\Omega_{\mu}^{\rho\sigma}e_{\rho}^{b}e_{\sigma}^{c},
\end{eqnarray}
where $\Omega_{\mu}^{\rho\sigma}$ is symmetrical about $\rho$ and $\sigma$, and it also satisfies the traceless condition $\Omega_{\mu}^{\rho\sigma}g_{\rho\sigma}=0$. From Eq.~(\ref{sec2-cs-tor-assm-1}), we can obtain the $SL(2,R)$ connection $\omega_{\mu}^{a}$
 \begin{eqnarray}
\label{sec2-cs-tor-assm-1-sol}
\omega_{\mu}^{a}=\frac{1}{2}\epsilon_{\hspace{2mm}c}^{ab}
E^{\sigma}_{b}(\partial_{\mu}e_{\sigma}^{c}-\Gamma^{\rho}_{\mu\sigma}e_{\rho}^{c}),
\end{eqnarray}
and Eq.~(\ref{sec2-cs-tor-assm-1}) also yields the metric compatibility condition
\begin{eqnarray}
\label{sec2-cs-tor-assm-mcom}
\partial_{\mu}g_{\alpha\beta}=\Gamma^{\rho}_{\mu\alpha}g_{\rho\beta}+\Gamma^{\rho}_{\mu\beta}g_{\rho\alpha},
\end{eqnarray}
which requires the connection to be
\begin{eqnarray}
\label{sec2-cs-tor-assm-mcom-con}
\Gamma^{\rho}_{\alpha\beta}&=&\bar{\Gamma}^{\rho}_{\alpha\beta}-g^{\rho\sigma}(T_{\alpha\sigma}^{\tau}g_{\tau\beta}
+T_{\beta\sigma}^{\tau}g_{\tau\alpha})+T_{\alpha\beta}^{\rho},\\
\label{sec2-cs-tor-assm-mcom-con-1}
\bar{\Gamma}^{\rho}_{\alpha\beta}&=&\frac{1}{2}g^{\rho\sigma}
(\partial_{\alpha}g_{\sigma\beta}+\partial_{\beta}g_{\sigma\alpha}-\partial_{\sigma}g_{\alpha\beta}),
\end{eqnarray}
where $T_{\alpha\beta}^{\rho}$ is the torsion tensor, which is antisymmetric about $\alpha$ and $\beta$. In terms of the variables in Eqs.~(\ref{sec2-lie-we-metric})-(\ref{sec2-lie-we-metric-anti}), (\ref{sec2-cs-tor-assm-2}) and~(\ref{sec2-cs-tor-assm-1-sol}), the action~(\ref{sec2-cs-lag-pal}) can be rewritten as
\begin{eqnarray}
\label{sec2-cs-lag-re-0}
S&=&\frac{1}{16\pi{G}}\int{d^3x}\sqrt{-g}\mathscr{L},\\
\mathscr{L}&=&\mathscr{L}_{1}+4\lambda^2(\mathscr{L}_{2}+\mathscr{L}_{3}+\mathscr{L}_{4}).\nonumber
\end{eqnarray}
In Eq.~(\ref{sec2-cs-lag-re-0}), $\mathscr{L}_{1}$ is
\begin{eqnarray}
\label{sec2-cs-lag-re-0a}
\mathscr{L}_{1}&=&R-\frac{2}{l^2},
\end{eqnarray}
where
\begin{eqnarray}
\label{sec2-cs-lag-re-1}
R^{\sigma}_{\hspace{1mm}\rho\mu\nu}=\partial_{\mu}\Gamma^{\sigma}_{\nu\rho}-\partial_{\nu}\Gamma^{\sigma}_{\mu\rho}
+\Gamma^{\sigma}_{\mu\tau}\Gamma^{\tau}_{\nu\rho}-\Gamma^{\sigma}_{\nu\tau}\Gamma^{\tau}_{\mu\rho}
\end{eqnarray}
is the Riemann curvature, and $R=g^{\alpha\beta}R^{\sigma}_{\hspace{1mm}\alpha\beta\sigma}$ is the Ricci scalar. $\mathscr{L}_{2}$, $\mathscr{L}_{3}$ and $\mathscr{L}_{4}$ are given by
\begin{subequations}
\begin{eqnarray}
\label{sec2-cs-lag-re-0b}
\mathscr{L}_{2}&=&g_{\alpha\beta}(\Omega^{\alpha\sigma}_{\rho}\Omega^{\beta\rho}_{\sigma}
-\Omega^{\alpha\sigma}_{\sigma}\Omega^{\beta\rho}_{\rho})\\
\label{sec2-cs-lag-re-0c}
\mathscr{L}_{3}&=&\frac{1}{l^2}
g_{\alpha\beta}(h^{\alpha\sigma}_{\rho}h^{\beta\rho}_{\sigma}
-h^{\alpha\sigma}_{\sigma}h^{\beta\rho}_{\rho}),\\
\label{sec2-cs-lag-re-0d}
\mathscr{L}_{4}&=&\varepsilon^{\mu\nu\alpha}(\nabla_{\mu}\Omega^{\rho\sigma}_{\nu}
+T^{\tau}_{\mu\nu}\Omega^{\rho\sigma}_{\tau})h_{\alpha\rho\sigma},
\end{eqnarray}
\end{subequations}
where
\begin{eqnarray}
\label{sec2-cs-lag-re-0d-def}
\nabla_{\mu}\Omega^{\alpha\beta}_{\nu}=\partial_{\mu}\Omega^{\alpha\beta}_{\nu}
-\Gamma^{\sigma}_{\mu\nu}\Omega^{\alpha\beta}_{\sigma}+\Gamma^{\alpha}_{\mu\sigma}\Omega^{\sigma\beta}_{\nu}
+\Gamma^{\beta}_{\mu\sigma}\Omega^{\alpha\sigma}_{\nu}
\end{eqnarray}
 is the covariant derivative associated with the connection $\Gamma^{\sigma}_{\mu\nu}$. $h^{\alpha\beta}_{\mu}=g^{\alpha\rho}g^{\beta\sigma}h_{\mu\rho\sigma}$, that is, we always lower and raise the indices through $g_{\alpha\beta}$ and its inverse $g^{\alpha\beta}$. From the above, we saw that $\mathscr{L}_{1}$ is the action of the conventional spin-2 gravity with the cosmological constant. $\mathscr{L}_{4}$ is a topologically likewise coupling term. The meaning of $\mathscr{L}_{2}$ would be clear if we know the expression of $\Omega_{\mu}^{\rho\sigma}$. Now the action~(\ref{sec2-cs-lag-re-0}) has a metric-like formulation, but it is a first-order action about $\Omega_{\mu}^{\alpha\beta}$ and $h_{\mu\alpha\beta}$. In order to obtain a second-order formulation, we need to solve the torsion constraints~(\ref{sec2-cs-lag-re-tor-1}) and~(\ref{sec2-cs-lag-re-tor-2}). The torsion constraint~(\ref{sec2-cs-lag-re-tor-1}) can be reformulated as
\begin{eqnarray}
\label{sec2-cs-tor-assm-re-3}
-T^{\gamma}_{\alpha\beta}=2\lambda^2
g_{\tau\mu}(\Omega_{\alpha}^{\sigma\tau}h_{\beta\sigma\rho}-
\Omega_{\beta}^{\sigma\tau}h_{\alpha\sigma\rho})\varepsilon^{\mu\rho\gamma},
\end{eqnarray}
and the torsion constraint~(\ref{sec2-cs-lag-re-tor-2}) can be reformulated as
\begin{subequations}
\begin{eqnarray}
\label{sec2-cs-tor-assm-re-1}
-K^{\gamma}_{\alpha\beta}&=&(\Omega_{\alpha}^{\gamma\tau}g_{\tau\beta}
-\Omega_{\rho}^{\rho\tau}g_{\tau\alpha}\delta^{\gamma}_{\beta})+(\alpha\leftrightarrow\beta),\\
\label{sec2-cs-tor-assm-re-2}
K^{\gamma}_{\alpha\beta}&=&\varepsilon^{\rho\sigma\gamma}(\nabla_{\rho}h_{\sigma\alpha\beta}
+T^{\tau}_{\rho\sigma}h_{\tau\alpha\beta}).
\end{eqnarray}
\end{subequations}
Eqs.~(\ref{sec2-cs-tor-assm-re-3}) and~(\ref{sec2-cs-tor-assm-re-1}) are derived from Eqs.~(\ref{sec2-cs-lag-re-tor-1}) and~(\ref{sec2-cs-lag-re-tor-2}) by multiplying the frame-like fields $E^{\alpha}_{a}$ or $e_{\beta}^{b}$. Alternatively, they can also be derived through variations of the action~(\ref{sec2-cs-lag-re-0}) regarding to $T^{\gamma}_{\alpha\beta}$ and $\Omega_{\mu}^{\rho\sigma}$ respectively.
Eqs.~(\ref{sec2-cs-tor-assm-re-3}) and~(\ref{sec2-cs-tor-assm-re-1}) are coupling equations about $T^{\gamma}_{\alpha\beta}$ and $\Omega_{\mu}^{\rho\sigma}$. Eq.~(\ref{sec2-cs-tor-assm-re-3}) demonstrates that the torsion is determined by the higher-spin fields, which provides the action~(\ref{sec2-cs-lag-re-0}) with the interpretation as a Einstein-Cartan-Sciama-Kibble theory~\cite{Hehl:1976kj}. The solution of Eq.~(\ref{sec2-cs-tor-assm-re-1}) can express the connection $\Omega_{\mu}^{\rho\sigma}$ with $h_{\mu\alpha\beta}$ and its derivatives. A solution of Eq.~(\ref{sec2-cs-tor-assm-re-1}) is
\begin{eqnarray}
\label{sec2-cs-tor-assm-re-1-sol}
\Omega_{\mu}^{\alpha\beta}&=&\frac{1}{2}\bigr(g^{\alpha\sigma}K^{\beta}_{\mu\sigma}+g^{\beta\sigma}K^{\alpha}_{\mu\sigma}
-\frac{2}{3}K^{\sigma}_{\mu\sigma}g^{\alpha\beta}\bigr)\\
&-&\frac{1}{2}g^{\alpha\rho}g^{\beta\sigma}g_{\mu\tau}K^{\tau}_{\rho\sigma}.\nonumber
\end{eqnarray}
Through this expression, $\Omega_{\mu}^{\alpha\beta}$ can be eliminated from Eqs.~(\ref{sec2-cs-lag-re-0b}) and~(\ref{sec2-cs-lag-re-0d}), and Eqs.~(\ref{sec2-cs-lag-re-0b}) and~(\ref{sec2-cs-lag-re-0d}) can be regarded as the kinetic terms of the spin-3 fields $h_{\mu\alpha\beta}$. Eq.~(\ref{sec2-cs-lag-re-0d}) looks like a Fierz-Pauli type massive term of $h_{\alpha\mu\nu}$. However, because the background solution of the action~(\ref{sec2-cs-lag-re-0}) is the anti-de Sitter space-time. Eq.~(\ref{sec2-cs-lag-re-0d}) plays the role to ensure the 3D diffeomorphism invariance of the action, but does not mean that the spin-3 field $h_{\mu\alpha\beta}$ is massive~\cite{Campoleoni:2012th,Deser:2001pe}. We can further attempt to solve the torsion constraints~(\ref{sec2-cs-tor-assm-re-3}). In 3D,  $T^{\gamma}_{\alpha\beta}$ is equivalent to a rank (2,0) tensor through the definition
\begin{eqnarray}
\label{sec2-cs-tor-re-def}
T^{\alpha\beta}=-\varepsilon^{\beta\rho\sigma}T^{\alpha}_{\rho\sigma},~~
T^{\gamma}_{\alpha\beta}=\frac{1}{2}T^{\gamma\rho}\varepsilon_{\rho\alpha\beta}.
\end{eqnarray}
Substituting $\Omega_{\mu}^{\alpha\beta}$ into Eq.~(\ref{sec2-cs-tor-assm-re-3}), we can obtain an equation of $T^{\alpha\beta}$
\begin{eqnarray}
\label{sec2-cs-tor-re-def-1}
T^{\alpha\beta}+4\lambda^2T^{\rho\sigma}M^{\alpha\beta}_{\rho\sigma}=4\lambda^2
\bar{\Omega}_{\theta}^{\sigma\tau}g_{\tau\mu}h_{\nu\sigma\rho}\varepsilon^{\mu\rho\alpha}\varepsilon^{\theta\nu\beta},
\end{eqnarray}
where $\bar{\Omega}_{\mu}^{\alpha\beta}$ is defined as $\Omega_{\mu}^{\alpha\beta}$ in~(\ref{sec2-cs-tor-assm-re-1-sol}) but with the connection $\Gamma^{\tau}_{\alpha\beta}$ replaced by the Levi-Civita connection $\bar{\Gamma}^{\tau}_{\alpha\beta}$. $M^{\alpha\beta}_{\rho\sigma}$ is a complicated algebraic function of $g_{\alpha\beta}$ and $h_{\rho\alpha\beta}$, which does not have a compact expression. To solve $T^{\alpha\beta}$, we need to know the inverse of $(\delta^{\alpha}_{\rho}\delta^{\beta}_{\sigma}+4\lambda^2M^{\alpha\beta}_{\rho\sigma})$, which is obtainable perturbatively or non-perturbatively in a algebraic way through the Caylay-Hamilton method. The first order approximation of  $T^{\alpha\beta}$ is given by the right side of Eq.~(\ref{sec2-cs-tor-re-def-1}). In this paper, we keep the torsion constraint~(\ref{sec2-cs-tor-assm-re-3}) intact in order that the action~(\ref{sec2-cs-lag-re-0}) has a concise formulation, then the action~(\ref{sec2-cs-lag-re-0}) is a ECSK theory coupled with the higher-spin fields $h_{\rho\alpha\beta}$.

\textit{Equations of motion}.{\textemdash}In order to obtain a transparent Lagrangian for $h_{\rho\alpha\beta}$, firstly we rewrite the Lagrangian $\mathscr{L}_{4}$ as
\begin{eqnarray}
\label{sec2-cs-lag-re-0d-1}
\mathscr{L}_{4}&=&\varepsilon^{\mu\nu\alpha}(\nabla_{\mu}h_{\nu\rho\sigma}
+T^{\tau}_{\mu\nu}h_{\tau\rho\sigma})\Omega_{\alpha}^{\rho\sigma}\\
&+&\frac{1}{\sqrt{-g}}\partial_{\mu}(\sqrt{-g}\varepsilon^{\mu\nu\alpha}\Omega_{\nu}^{\rho\sigma}h_{\alpha\rho\sigma}).\nonumber
\end{eqnarray}
The second line of this equation is a total divergence term. Substituting the solution~(\ref{sec2-cs-tor-assm-re-1-sol}) of $\Omega_{\alpha}^{\rho\sigma}$ into Eqs.~(\ref{sec2-cs-lag-re-0b}) and~(\ref{sec2-cs-lag-re-0d-1}), we obtain a new Lagrangian
\begin{eqnarray}
\label{sec2-cs-lag-re-0d-1a}
\mathscr{L}_{2}+\mathscr{L}_{4}&=&-\frac{1}{4}(g^{\mu\alpha}g^{\nu\beta}-g^{\mu\beta}g^{\nu\alpha})
\hat{\nabla}_{\mu}h_{\nu}^{\rho\sigma}\hat{\nabla}_{\alpha}h_{\beta\rho\sigma}\nonumber\\
&-&\frac{1}{2}g^{\tau\theta}\varepsilon^{\mu\nu\rho}\varepsilon^{\alpha\beta\sigma}
\hat{\nabla}_{\mu}h_{\nu\sigma\tau}\hat{\nabla}_{\alpha}h_{\beta\rho\theta},
\end{eqnarray}
where we have use $\hat{\nabla}_{\mu}h_{\nu\rho\sigma}=\nabla_{\mu}h_{\nu\rho\sigma}+T^{\tau}_{\mu\nu}h_{\tau\rho\sigma}$ to achieve a compact expression, and the divergence term in Eq.~(\ref{sec2-cs-lag-re-0d-1}) was omitted. This Lagrangian has the Maxwell-like formulation, which is quadratic about the field strength. The identity
\begin{eqnarray}
\label{sec2-cs-lag-re-0d-1a-id}
\varepsilon^{\mu\nu\rho}\varepsilon^{\alpha\beta\sigma}&=&g^{\mu\alpha}g^{\nu\sigma}g^{\rho\beta}
+(g^{\nu\alpha}g^{\mu\beta}-g^{\mu\alpha}g^{\nu\beta})g^{\rho\sigma}\\
&-&g^{\nu\alpha}g^{\mu\sigma}g^{\rho\beta}+(g^{\mu\sigma}g^{\nu\beta}-g^{\nu\sigma}g^{\mu\beta})g^{\rho\alpha}\nonumber
\end{eqnarray}
can be further used to rewrite Eq.~(\ref{sec2-cs-lag-re-0d-1a-id}) into a conventional formulation. Now we discuss the equations of motion about $g_{\alpha\beta}$ and $h_{\rho\alpha\beta}$. Their equations of motion are given by the zero curvature condition~(\ref{sec2-cs-lag-pal-eom}), which can be decomposed into two equations as the torsion constraints~(\ref{sec2-cs-lag-re-tor-1}) and~(\ref{sec2-cs-lag-re-tor-2}). Firstly, from Eq.~(\ref{sec2-cs-lag-pal-eom}), we can obtain
\begin{subequations}
\begin{eqnarray}
\label{sec2-cs-lag-pal-eom-re-1a}
2\lambda^2\mathcal{T}_{\mu\nu}&=&R_{\mu\nu}-\frac{1}{2}{R}g_{\mu\nu}+\frac{1}{l^2}g_{\mu\nu},\\
\label{sec2-cs-lag-pal-eom-re-1b}
\mathcal{T}_{\mu\nu}&=&\mathscr{L}_{2}g_{\mu\nu}
+2(\Omega_{\tau}^{\sigma\tau}\Omega_{\nu\sigma\mu}-\Omega_{\nu}^{\sigma\tau}\Omega_{\tau\sigma\mu})\\
&+&\mathscr{L}_{3}g_{\mu\nu}+\frac{2}{l^2}(h_{\tau}^{\sigma\tau}h_{\nu\sigma\mu}-h_{\nu}^{\sigma\tau}h_{\tau\sigma\mu}).\nonumber
\end{eqnarray}
\end{subequations}
Here $R_{\mu\nu}=R^{\sigma}_{\hspace{1mm}\mu\nu\sigma}$ is the Ricci tensor. Eq.~(\ref{sec2-cs-lag-pal-eom-re-1a}) is the equation of motion of the spin-2 field $g_{\mu\nu}$, which has the same formulation with the Einstein equation. Because the connection has a torsion, $R_{\mu\nu}$ is not symmetric about its indices~\cite{Hehl:1976kj}.  $\mathcal{T}_{\mu\nu}$ is the energy-momentum tensor contributed by the higher spin fields, and it is also not symmetric. From Eq.~(\ref{sec2-cs-lag-pal-eom}), we can also obtain
\begin{subequations}
\begin{eqnarray}
\label{sec2-cs-lag-pal-eom-re-2a}
-H^{\gamma}_{\hspace{1mm}\alpha\beta}&=&\frac{1}{l^2}(h_{\alpha}^{\gamma\tau}g_{\tau\beta}
-h_{\rho}^{\rho\tau}g_{\tau\alpha}\delta^{\gamma}_{\beta})
+(\alpha\leftrightarrow\beta),\\
\label{sec2-cs-lag-pal-eom-re-2b}
H^{\gamma}_{\alpha\beta}&=&\varepsilon^{\rho\sigma\gamma}(\nabla_{\rho}\Omega_{\sigma\alpha\beta}
+T^{\tau}_{\rho\sigma}\Omega_{\tau\alpha\beta}).
\end{eqnarray}
\end{subequations}
Substituting the solution~(\ref{sec2-cs-tor-assm-re-1-sol}) of $\Omega_{\mu}^{\alpha\beta}$ into~(\ref{sec2-cs-lag-pal-eom-re-2b}),  we can obtain the equations of motion about $h_{\mu\alpha\beta}$, though they do not have a compact expression as the action~(\ref{sec2-cs-lag-re-0d-1a}). We saw that Eqs.~(\ref{sec2-cs-lag-pal-eom-re-2a}) and~(\ref{sec2-cs-lag-pal-eom-re-2b}) have the same structure as Eqs.~(\ref{sec2-cs-tor-assm-re-1}) and~(\ref{sec2-cs-tor-assm-re-2}). From Eq.~(\ref{sec2-cs-lag-pal-eom-re-2a}), we have
\begin{eqnarray}
\label{sec2-cs-tor-assm-re-2-sol-h}
\frac{1}{l^2}h_{\mu\alpha\beta}&=&\frac{1}{2}\bigr(g_{\alpha\sigma}H^{\sigma}_{\mu\beta}+g_{\beta\sigma}H^{\sigma}_{\mu\alpha}
-\frac{2}{3}g_{\alpha\beta}H^{\sigma}_{\mu\sigma}\bigr)\\
&-&\frac{1}{2}g_{\mu\tau}H^{\tau}_{\alpha\beta}.\nonumber
\end{eqnarray}
which is an equivalent formulation of Eq.~(\ref{sec2-cs-lag-pal-eom-re-2a}), and it is similar to Eq.~(\ref{sec2-cs-tor-assm-re-1-sol}).

\textit{Duality-like Transformation}.{\textemdash}We have noticed that the similarity between Eq.~(\ref{sec2-cs-tor-assm-re-2-sol-h}) and Eq.~(\ref{sec2-cs-tor-assm-re-1-sol}), which indicate a duality-like transformation between $\Omega_{\mu}^{\alpha\beta}$ and $h_{\mu}^{\alpha\beta}$. To make this transformation transparent, we rewrite the Lagrangian $\mathscr{L}_{4}$ as
\begin{eqnarray}
\label{sec2-cs-lag-re-0d-re-2}
\mathscr{L}_{4}&=&\frac{1}{2}\varepsilon^{\mu\nu\alpha}(\nabla_{\mu}\Omega^{\rho\sigma}_{\nu}
+T^{\tau}_{\mu\nu}\Omega^{\rho\sigma}_{\tau})h_{\alpha\rho\sigma}\\
&+&\frac{1}{2}\varepsilon^{\mu\nu\alpha}(\nabla_{\mu}h_{\nu\rho\sigma}
+T^{\tau}_{\mu\nu}h_{\tau\rho\sigma})\Omega^{\rho\sigma}_{\alpha}\nonumber\\
&+&\frac{1}{2}\frac{1}{\sqrt{-g}}\partial_{\mu}(\sqrt{-g}\varepsilon^{\mu\nu\alpha}\Omega_{\nu}^{\rho\sigma}h_{\alpha\rho\sigma}).\nonumber
\end{eqnarray}
If we do not consider the divergence term in Eq.~(\ref{sec2-cs-lag-re-0d-re-2}), then the action~(\ref{sec2-cs-lag-re-0}) is invariant under the duality likewise transformation
\begin{eqnarray}
\label{sec2-cs-lag-re-0d-re-a}
\tilde{\Omega}^{\rho\sigma}_{\mu}=\frac{1}{l}h^{\rho\sigma}_{\mu},~~
\frac{1}{l}\tilde{h}^{\rho\sigma}_{\mu}=\Omega^{\rho\sigma}_{\mu}.
\end{eqnarray}
We can furthure define
\begin{subequations}
\begin{eqnarray}
\label{sec2-cs-lag-re-0d-de-a}
\Omega^{\hspace{1mm}\rho\sigma}_{\mu}&=&\frac{1}{\sqrt{2}}(U^{\rho\sigma}_{\mu}-V^{\rho\sigma}_{\mu}),\\
\label{sec2-cs-lag-re-0d-de-b}
\frac{1}{l}{h}^{\rho\sigma}_{\mu}&=&
\frac{1}{\sqrt{2}}(U^{\rho\sigma}_{\mu}+V^{\rho\sigma}_{\mu}),
\end{eqnarray}
\end{subequations}
then $\mathscr{L}_{2}$ and $\mathscr{L}_{3}$ can be rewritten as
\begin{eqnarray}
\label{sec3-cs-lag-re-0d-re-a}
\mathscr{L}_{2}&+&\mathscr{L}_{3}=g_{\alpha\beta}(U^{\alpha\sigma}_{\rho}U^{\beta\rho}_{\sigma}
-U^{\alpha\sigma}_{\sigma}U^{\beta\rho}_{\rho})\\
&+&g_{\alpha\beta}(V^{\alpha\sigma}_{\rho}V^{\beta\rho}_{\sigma}
-V^{\alpha\sigma}_{\sigma}V^{\beta\rho}_{\rho}),\nonumber
\end{eqnarray}
and $\mathscr{L}_{4}$ can be rewritten as
\begin{eqnarray}
\label{sec3-cs-lag-re-0d-re-b}
\mathscr{L}_{4}&=&\frac{l}{2}\varepsilon^{\mu\nu\alpha}(\nabla_{\mu}U^{\rho\sigma}_{\nu}
+T^{\tau}_{\mu\nu}U^{\rho\sigma}_{\tau})U_{\alpha\rho\sigma}\\
&-&\frac{l}{2}\varepsilon^{\mu\nu\alpha}(\nabla_{\mu}V_{\nu\rho\sigma}
+T^{\tau}_{\mu\nu}V_{\tau\rho\sigma})V^{\rho\sigma}_{\alpha}\nonumber
\end{eqnarray}
up to the divergence term in Eq.~(\ref{sec2-cs-lag-re-0d-re-2}). The torsion constraint~(\ref{sec2-cs-tor-assm-re-3}) can be rewritten as
\begin{eqnarray}
\label{sec2-cs-tor-assm-re-1-uv}
-T^{\gamma}_{\alpha\beta}=2\lambda^2
g_{\tau\mu}(U_{\alpha}^{\sigma\tau}U_{\beta\sigma\rho}-
V_{\beta}^{\sigma\tau}V_{\alpha\sigma\rho})\varepsilon^{\mu\rho\gamma}.
\end{eqnarray}
From the above, we saw that the cross products of $U^{\alpha\beta}_{\mu}$ and $V^{\alpha\beta}_{\mu}$ are eliminated from Eqs.~(\ref{sec3-cs-lag-re-0d-re-a}) and~(\ref{sec3-cs-lag-re-0d-re-b}). So the action~(\ref{sec2-cs-lag-re-0}) can be interpreted as the spin-2 gravity interacting with two rank~(2,1)~tensor fields. However, $U^{\alpha\beta}_{\mu}$ and $V^{\alpha\beta}_{\mu}$ are not free fields, and their interactions is provided by the torsion constraint~(\ref{sec2-cs-tor-assm-re-1-uv}) through the covariant direvative. From Eqs.~(\ref{sec2-cs-tor-assm-re-1}) and~(\ref{sec2-cs-lag-pal-eom-re-2a}), we have
\begin{subequations}
\begin{eqnarray}
\label{sec2-cs-lag-pal-eom-re-2a-u}
-\tilde{H}^{\gamma}_{\hspace{1mm}\alpha\beta}&=&\frac{1}{l}(U_{\alpha}^{\gamma\tau}g_{\tau\beta}
-U_{\rho}^{\rho\tau}g_{\tau\alpha}\delta^{\gamma}_{\beta})
+(\alpha\leftrightarrow\beta),\\
\label{sec2-cs-lag-pal-eom-re-2b-u}
\tilde{H}^{\gamma}_{\alpha\beta}&=&\varepsilon^{\rho\sigma\gamma}(\nabla_{\rho}U_{\sigma\alpha\beta}
+T^{\tau}_{\rho\sigma}U_{\tau\alpha\beta}).
\end{eqnarray}
\end{subequations}
If we omit the torsion constraint~(\ref{sec2-cs-tor-assm-re-1-uv}), then Eq.~(\ref{sec2-cs-lag-pal-eom-re-2a-u}) is linear about $U_{\alpha}^{\beta\gamma}$. $\tilde{H}^{\gamma}_{\alpha\beta}$ can be interpreted as the field which is dual to the Maxwell-like field strength
\begin{eqnarray}
\label{sec2-cs-lag-pal-eom-re-2a-u-1}
\mathcal{F}_{\mu\nu\alpha\beta}=\nabla_{\mu}U_{\nu\alpha\beta}-\nabla_{\nu}U_{\mu\alpha\beta}.
\end{eqnarray}
While the right side of Eq.~(\ref{sec2-cs-lag-pal-eom-re-2a-u}) is a part of $U_{\alpha\beta\gamma}$ which is symmetric and traceless about its first two indices. So Eq.~(\ref{sec2-cs-lag-pal-eom-re-2a-u}) has the meaning that the dual of the field strength of $U_{\alpha\beta\gamma}$ is the minus of its symmetric and traceless part about its first two indices. Similarly, from Eqs.~(\ref{sec2-cs-tor-assm-re-1}) and~(\ref{sec2-cs-lag-pal-eom-re-2a}), we also have
\begin{subequations}
\begin{eqnarray}
\label{sec2-cs-lag-pal-eom-re-2a-v}
\tilde{K}^{\gamma}_{\hspace{1mm}\alpha\beta}&=&\frac{1}{l}(V_{\alpha}^{\gamma\tau}g_{\tau\beta}
-V_{\rho}^{\rho\tau}g_{\tau\alpha}\delta^{\gamma}_{\beta})
+(\alpha\leftrightarrow\beta),\\
\label{sec2-cs-lag-pal-eom-re-2b-v}
\tilde{K}^{\gamma}_{\alpha\beta}&=&\varepsilon^{\rho\sigma\gamma}(\nabla_{\rho}V_{\sigma\alpha\beta}
+T^{\tau}_{\rho\sigma}V_{\tau\alpha\beta}),
\end{eqnarray}
\end{subequations}
which have the interpretations similar to Eqs.~(\ref{sec2-cs-lag-pal-eom-re-2a-u}) and~(\ref{sec2-cs-lag-pal-eom-re-2b-u}).

\textit{Generalized diffeomorphism}.{\textemdash}Now we discuss the potential symmetries of the action~(\ref{sec2-cs-lag-re-0}). These symmetries can be induced from the symmetries of the Chern-Simons action~(\ref{sec2-cs-lag}). If the boundary terms are negligible, then the CS action~(\ref{sec2-cs-lag}) is invariant under the infinitesimal $SL(3,R){\times}SL(3,R)$ gauge transformations
\begin{subequations}
\begin{eqnarray}
\label{sec2-cs-gtr-a}
\delta{A}&=&d\zeta+[A,\zeta],\\
\label{sec2-cs-gtr-b}
\delta{\bar{A}}&=&d\bar{\zeta}+[\bar{A},\bar{\zeta}].
\end{eqnarray}
\end{subequations}
In terms of the decomposition~(\ref{sec2-cs-frame}), we have
\begin{eqnarray}
\label{sec2-cs-gtr-lor-a}
\delta{\omega}&=&d\Lambda+[\omega,\Lambda]+\frac{1}{l}[e,\xi],\\
\label{sec2-cs-gtr-diff-b}
\frac{1}{l}\delta{e}&=&d\xi+[\omega,\xi]+\frac{1}{l}[e,\Lambda],
\end{eqnarray}
where we have defined
\begin{eqnarray}
\label{sec2-cs-gtr-diff-a-de}
\xi=\frac{1}{2}(\zeta-\bar{\zeta}),~~\Lambda=\frac{1}{2}(\zeta+\bar{\zeta}).
\end{eqnarray}
 Now we focus on the transformation~(\ref{sec2-cs-gtr-diff-b}). $\Lambda$ and $\xi$ have the $SL(2,R)$ decomposition
 \begin{eqnarray}
\label{sec2-cs-gtr-diff-a-de-lx}
\Lambda=\Lambda^{a}J_a+\Lambda^{bc}Q_{bc},~~\xi=\xi^{a}J_a+\xi^{bc}Q_{bc},
\end{eqnarray}
 where $\Lambda^{ab}$ and  $\xi^{ab}$ are symmetrical and traceless. If $\xi=0$,  then Eq.~(\ref{sec2-cs-gtr-diff-b}) yields the local Lorentz transformations
 \begin{subequations}
\begin{eqnarray}
\label{sec2-cs-gtr-lor-a-re}
\delta_{\Lambda}{g_{\mu\nu}}&=&
4\lambda^2\left(h_{\mu}^{\rho\sigma}\Lambda_{\rho\tau}g^{\tau\theta}\varepsilon_{\sigma\theta\nu}+(\mu\leftrightarrow\nu)\right) ,\\
\label{sec2-cs-gtr-lor-b-re}
\delta_{\Lambda}{h_{\alpha\mu\nu}}&=&\left(\Lambda_{\mu\rho}g^{\rho\theta}\varepsilon_{\alpha\theta\nu}+(\mu\leftrightarrow\nu)\right)\\
&+&4\lambda^2\left({h}_{\alpha}^{\rho\sigma}{h}_{\mu}^{\tau\theta}
g_{\rho\nu}\Lambda_{\tau\beta}g^{\beta\gamma}\varepsilon_{\theta\sigma\gamma}+(\mu\leftrightarrow\nu)\right),\nonumber
\end{eqnarray}
\end{subequations}
where $\Lambda_{\mu\nu}=\Lambda_{ab}e^{a}_{\mu}e^{b}_{\nu}$ is symmetrical and satisfies the traceless condition $g^{\mu\nu}\Lambda_{\mu\nu}=0$. We saw that the local lorentz transformations only depend on the parameter $\Lambda_{\rho\sigma}$, but it is independent of $\Lambda_{\mu}=\Lambda_{a}e^{a}_{\mu}$. This is consistent with the fact that $g_{\mu\nu}$ and $h_{\alpha\mu\nu}$ are $SL(2,R)$ invariant variables, but they are not $SL(3,R)$ invariant ones. Otherwise, If $\Lambda=0$, from Eq.~(\ref{sec2-cs-gtr-diff-b}), we can obtain the generalized diffeomorphism
\begin{subequations}
\begin{eqnarray}
\label{sec2-cs-gtr-diff-a-re}
\frac{1}{l}\delta_{\xi}{g_{\mu\nu}}&=&\nabla_{\mu}\xi_{\nu}+\nabla_{\nu}\xi_{\mu}\\
&+&4\lambda^2\left(\Omega_{\mu}^{\tau\beta}\xi_{\tau\rho}g^{\rho\sigma}\varepsilon_{\beta\sigma\nu}
+(\mu\leftrightarrow\nu)\right) ,\nonumber\\
\label{sec2-cs-gtr-diff-b-re}
\frac{1}{l}\delta_{\xi}{h_{\alpha\mu\nu}}&=&\nabla_{\alpha}\xi_{\mu\nu}+\left(h_{\alpha\rho\mu}g^{\rho\gamma}
\nabla_{\nu}\xi_{\gamma}+(\mu\leftrightarrow\nu)\right)\\
&+&4\lambda^2\left(h_{\alpha\rho\mu}g^{\rho\gamma}
\Omega_{\nu}^{\tau\beta}\xi_{\tau\theta}g^{\theta\sigma}\varepsilon_{\beta\sigma\gamma}
+(\mu\leftrightarrow\nu)\right),\nonumber
\end{eqnarray}
\end{subequations}
where $\nabla_{\mu}\xi_{\nu}=\partial_{\mu}\xi_{\nu}-\Gamma^{\rho}_{\mu\nu}\xi_{\rho}$ is the covariant derivative with the connection~(\ref{sec2-cs-tor-assm-mcom-con}). $\xi_{\mu}=\xi_{a}e^{a}_{\mu}$, and $\xi_{\mu\nu}=\xi_{ab}e^{a}_{\mu}e^{b}_{\nu}$ is symmetrical and traceless. $\Omega_{\nu}^{\tau\beta}$ is defined by Eq.~(\ref{sec2-cs-tor-assm-re-1-sol}). If $h_{\alpha\mu\nu}$ is small, then the second term of the right side of Eq.~(\ref{sec2-cs-gtr-diff-a-re}) is negligible. Eq.~(\ref{sec2-cs-gtr-diff-a-re}) yields the conventional diffeomorphism for the spin-2 gravity.

\textit{Conclusions}.{\textemdash}We have provided a metric-like formulation for the $SL(3,R){\times}SL(3,R)$ Chern-Simons theory using the $SL(2,R)$ invariant variables. This metric-like formulation can be interpreted as a Einstein-Cartan-Sciama-Kibble theory~\cite{Hehl:1976kj}, in which the torsion is determined by the higher-spin fields. The local Lorentz transformation and the generalized diffeomorphism can be expressed with these metric-like fields manifestly. We also identify a duality-like transformation in this metric-like formulation. Because the Lie algebra of  $SL(N,R)$ has the decomposition under its sub algebra $SL(2,R)$ similar to that of $SL(3,R)$, the $SL(2,R)$ variables used here could also be useful to find a metric-like formulation for the $SL(N,R){\times}SL(N,R)$ Chern-Simons theory~\cite{Campoleoni:2010zq}, and the duality-like transformation discussed here could also be found in those theories.

{\textit {Acknowledgments}}.{\textemdash}This work was supported in part by Fondecyt~(Chile) grant 1100287 and by Project Basal under Contract No.~FB0821.

\bibliographystyle{apsrev4-1}
\bibliography{spinRef}

\end{document}